# A 3D index for measuring economic resilience with application to the modern international and global financial crises

**Dimitrios Tsiotas**

Department of Regional and Economic Development, Agricultural University of Athens, Nea Poli, Amfissa, 33100, Greece.

E-mail: tsiotas@aua.gr

**Abstract**

The study and measurement of economic resilience is ruled by high level of complexity related to the diverse structure, functionality, spatiality, and dynamics describing economic systems. Towards serving the demand of integration, this paper develops a three-dimensional (3D) index, capturing engineering, ecological, and evolutionary aspects of economic resilience that are considered separately in the current literature. The proposed index is computed on GDP data of worldwide countries, for the period 1960-2020, concerning 14 crises considered as shocks, and was found well defined in a conceptual context of its components. Its application on real-world data allows introducing a novel classification of countries in terms of economic resilience, and reveals geographical patterns and structural determinants of this attribute. Impressively enough, economic resilience appears positively related to major productivity coefficients, gravitationally driven, and depended on agricultural specialization, with high structural heterogeneity in the low class. Also, the analysis fills the literature gap by shaping the worldwide map of economic resilience, revealing geographical duality and centrifugal patterns in its geographical distribution, a relationship between diachronically good performance in economic resilience and geographical distance from the shocks' origin, and a continent differentiation expressed by the specialization of America in engineering resilience, Africa and Asia in ecological and evolutionary resilience, and a relative lag of Europe and Oceania. Finally, the analysis provides insights into the effect of the 2008 on the globe and supports a further research hypothesis that political instability is a main determinant of low economic resilience, addressing avenues of further research.

## 1. INTRODUCTION

Studying economic systems is a demanding process ruled by a high level of complexity (Metcalfe and Foster, 2007; Cristelli et al., 2015). Amongst the many that can be detected in the literature, one major aspect composing such complexity is ontological and regards the structure of a system, and particularly the elements that it consists of (people, labor, firms, banks, etc.), along with their attributes, scale (microeconomic, macroeconomic), space of embedding (topological, social, metric, etc.), the structural setting, and potential to connect (Whiteley, 2000; Granovetter, 2005; Boschma and Frenken, 2009; Potts, 2010). Another major aspect is industrial or sectorial (Dasgupta and Stiglitz, 1980; Combes, 2000; Zeira and Zoabi, 2015) and regards the system's economic functionality, such as trade, transportation, communication, information, recreation, etc. Another major aspect is geographical (Reggiani et al., 2002; Behrens and Thisse, 2007; Gluckler, 2007) and concerns the system's spatial (metric) configuration, such as its embedding in the geographical space, the geographical scale (punctual, local, urban, regional, national, international, and global), etc. A final major aspect is dynamic (Pasinetti, 1993; North, 1994; Brun et l., 2005; Kalecki, 2013) and concerns the system's evolution along the temporal dimension and its type (inclining, constant, or declining). This setting of complexity has made the economics discipline grow and develop many fields and subspecialties, attracting scholars of diverse backgrounds, producing multidisciplinary research. The multidisciplinary conceptualization of modern economics has become very fertile, offering the potential to study economic systems within a diverse context and under a high level of resolution.

Enjoying such multidisciplinary interest, recent research (Simmie and Martin, 2010; Lee, 2014; Martin and Sunley, 2014; Kitsos and Bishop, 2018; Buheji, 2019) has brought into the light the new concept of economic resilience, which is related to the capability of an economic system to respond to disturbances (shocks) and is defined about a shock, in terms of either recovering to the previous state of functionality or moving into a new one. The concept of resilience originates from biological, social, and engineering disciplines, such as ecology, biology, sociology, psychology, and physics (mechanics), but has become broad and integrated within the context of economics, economic geography, and regional science (Martin and Sunley, 2014; Modica and Reggiani, 2015). As expected, economic resilience is currently conceived within a richly diverse context (Simmie and Martin, 2010; Andreoni and Duriavig, 2013; Bristow et al., 2014), inheriting the complexity ruling relevant research and the disciplinary background of the researchers studying in this concept. For instance, when the system under study is an economic sector, then resilience has a sectorial configuration and is conceived as structural or sectorial resilience (Morkunas et al., 2018; Michel-Villarreal et al., 2019; Volkov et al., 2021). On the other hand, when a system enjoys a specific geographical configuration or reference (market area, city, region, etc.), resilience is defined in a regional context and is as regional resilience (Brakman et al., 2014; Capello et al., 2015; Giannakis and Bruggeman, 2017). Also, the engineering ability of a system to recover to its previous state of functionality defines the engineering resilience (Bristow et al., 2014; Capello et al., 2015; Buheji, 2019), whereas the ecological process of moving to a new state of functionality defines ecological resilience (Andreoni and Duriavig, 2013; Bristow et al., 2014; Kitsos and Bishop, 2018), and the dynamic ability to timely adjust to disturbances defines the evolutionary resilience (Martin and Sunley, 2014). The broad term of economic resilience highlights the economic configuration of a system under examination (Martin and Sunley, 2014), and its measurement is also described by a high level of complexity. As far as measurement is concerned, a variety of variables, indices, indicators, statistics, and models are available in the literature (Sensier et al., 2016; Kitsos and Bishop, 2018; Martin and Gardiner, 2018), quantitatively determining aspects of the multifaceted context of resilience. This broad availability of methods provides an excellent context for integrating the diverse temporal, spatial, and structural dimensions and for promoting thus a holistic conceptualization of resilience. However, current studies going beyond mono-variable considerations of economic resilience are constrained (Schwarz et al., 2011; Palaskas et al., 2015; Sdrolias et al., 2022), and they either build on proportion and pairwise indicators, or multivariable modeling approximating resilience as a single response variable, or on multilevel modeling that is usually counterintuitive. Moreover, although resilience has been thoroughly studied in the current literature at the international level (Brakman et al., 2014; Sensier et al., 2016; Xiao et al., 2017; Giannakis and Bruggeman, 2020) or the interregional scale at the national level (Kitsos and Bishop, 2018; Michel-Villarreal et al., 2019; Volkov et al., 2021), this property has not been examined in a comprehensive context globally (van Bergeijk et al., 2017; Martin et al., 2016). Towards serving the double demand of developing an integrated approach and at the same time providing an empirical study contributing to the dialogue at the global scale, this paper develops a three-dimensional (3D) index of economic resilience computed on 200 countries worldwide and empirically evaluates its reliability in the context of regional science. The proposed 3D index consists of components expressing diverse intrinsic properties (engineering, ecological, and evolutionary) of resilience and is computed on spatio-temporal GDP data of the period 1960-2020, for the available countries, about 15 major modern international and global financial crises. The results are evaluated for their determination ability and realism in the worldwide context of regional science and economics. The remainder of this paper is structured as follows: Section 2 provides a literature review; Section 3 presents the methodological framework of the proposed 3d index and the data of the study; Section 4 shows the results of the analysis and discusses them within the context of regional science and economics, and finally, in Section 5, conclusions are given.

## 2. LITERATURE REVIEW

Economic resilience is a concept related to the complex configuration of economic systems (Lee, 2014; Kitsos and Bishop, 2018) and their capability to respond to disturbances (shocks). Although resilience originates from engineering and biological disciplines, it has been flourished and transformed to economic resilience in the context of economics and applied geography (Martin and Sunley, 2014), which study economic systems in a composite spatio-temporal and structural context. Conceptually, resilience consists of five main pillars that are necessary to comprehend its nature as spherically as possible (Martin and Sunley, 2014): the first is related to the origin of a disturbance (shock), while the others relate to aspects of the system's behavior, such as its sensitivity (vulnerability) to the disturbance; its initial response (resistance) the shock; its adaptability to the disturbance (robustness); and its ability to recover (recoverability) or shift into another state of functionality. In current literature, economic resilience enjoys two main research strands (Martin and Sunley 2014; Kitsos and Bishop, 2018), the equilibrium and evolutionary approaches. The first strand conceptualizes resilience in an equilibrium context, where the response to a shock either follows an engineering process (Capello et al., 2015; Buheji, 2019), recovering the system into its pre-existing equilibrium (engineering resilience), or an ecological process (Bristow et al., 2014; Kitsos and Bishop, 2018), shifting the system towards a new level of functionality (ecological resilience). The second strand conceives resilience in a dynamic context, in the run of a system's adaptation to continuously changing conditions (Martin and Sunley, 2014). Another important aspect for conceiving resilience is the emergence of the shock, defining two periods in the time range of the system's evolution, the pre-shock and the post-shock period, and another two stages in the post-shock sequence of the system's response (Kitsos and Bishop, 2018), the recession (impact, downturn) and the recovery (rebound, adaptation) stage. The relativity of resilience to the shock highlights the importance of setting a meaningful reference in the definition and measurement of this concept (Martin and Sunley, 2014). Also, the geographical configuration adds to the relativity in the resilience's definition (Reggiani et al., 2002; Sensier et al., 2016; Kitsos and Bishop, 2018), since a system's resilience may differ when it is studied as an isolated spatial unit or as a part of spatial unity. In this debate, Kitsos and Bishop (2018) note that an interregional setting (such as considering subnational differences) in the study of resilience is more effective to drive better-targeted policies.

The by default reference of economic systems to a geographic framework made economic resilience a prominent topic in the research agenda of regional and urban science and applied geography, resulting in a conceptual transformation known as regional resilience (Psycharis et al., 2014; Giannakis and Bruggeman, 2017). In current literature, relevant research in regional resilience is rich and diverse. In terms of geographical scale, we can find at the intra-national level relevant studies, such as of: Schwarz et al. (2011), who examined the vulnerability and resilience of coastal communities in Solomon Islands in terms of social processes and coping with change; Andreoni and Duriavig (2013), who studied the resilience of municipalities in the south-east Bahia (Brazil) to the cocoa crisis of the 1990s in terms of land use diversification; Psycharis et al. (2014), who assessed the resilience of NUTS II and NUTS III regions in Greece to the 2008 economic crisis in the context of sectorial structure; Palaskas et al. (2015), who examined the resilience of Greek urban economies to the impact of the 2008 economic crisis in terms of sectorial heterogeneity; Martin et al. (2016), who studied the resilience of UK regions to four major recessions of the past half century in the context of shifting between economic cycles and competitiveness; Martin and Gardiner (2018), who examined the resilience of British cities to the last four major economic shocks of the period 1971-2015 in terms of the relation between resistance and recovery of shocks; Kitsos et al. (2019), who studied the economic resilience of local industrial embeddedness of UK regions (NUTS II) to the 2008 financial crisis in the context of their regional socioeconomic and demographic configuration; and many others. At the international level, we can find studies, such as of: Brakman et al. (2014), who examined the resilience of the EU regions to the 2008 economic crisis in terms of regional urbanization and specialization; Capello et al. (2015), who studied the regional resilience cities in Europe to the twenty-year period 1990–2011 in the context of production quality, external connectivity in cooperation networks, and urban infrastructure quality; Sensier et al. (2016), who examined regional economic resilience of European regions to several major economic shocks since the early 1990s in the context of resilience interconnection between economic shocks; Xiao et al. (2017), who studied regional resilience of European regions' economies to the 2008 economic crisis, in terms of their ability to develop new post-crisis industrial paths; Kitsos and Bishop (2018), who studied regional resilience of Great Britain's Local Authority Districts to 2008 crisis, in terms of the initial economic conditions, human capital, age structure, urbanization, and geography of regions; Giannakis and Papadas (2021), who studied regional resilience of 25 EU countries to the 2008 economic crisis in terms of spatial connectivity; and many others. As is evident, the case of EU prevails in the examination of regional resilience at the international level, perhaps due to the economic role, the institutional uniqueness in current history, and the size of national participation in the configuration of EU. The strand of studies conceiving resilience in a regional context highlights the importance and relativity of the spatial configuration in the resilience of regional economic systems. In this framework, the structural (sectorial) configuration of economic systems adds a further relativity degree. In particular, focusing on the sectorial structure instead of the geographical configuration of an economy has led to the emergence of industrial (or structural or sectorial) resilience (Sdrolias et al., 2022), which describes the elasticity of an economic sector against disturbances. Studies in this research strand (Morkunas et al., 2018; Michel-Villarreal et al., 2019; Papadopoulos et al., 2019; Volkov et al., 2021; Pnevmatikos et al., 2019) examine the resilience of an economic sector or activity either individually or compared to other sectors. This approach is of great importance because a resilient sector may provide a way to economic growth and development in periods of recession and economic decline (Pnevmatikos et al., 2019; Volkov et al. 2021).

The multidisciplinary framework of research makes the modeling and measurement of resilience a complicated task. Martin and Sunley (2014) observe a lack of a commonly accepted methodology in the empirical measurement of resilience and relatively little discussion of how it relates to regional development, competitiveness, or path dependence. For instance, engineering resilience is typically measured in terms of recovering speed to the equilibrium, whilst ecological by the force caused a system's shift to a new state of functionality (Modica and Reggiani, 2015; Kitsos and Bishop, 2018).

According to Martin and Sunley (2014), modeling approaches measuring economic resilience can divide into (*i*) qualitative case studies based on simple descriptive data, interviews with key actors, and interrogation of policies; (*ii*) quantitative resilience indices constructed on singular, composite, and comparative measures of (relative) resistance and recovery; (*iii*) time series models studying systems' evolution and duration to the shock's impact; and (*iv*) causal structural models, embedding resilience in regional economic models and examining counterfactual scenarios in absence of a shock. A common underlying component amongst these diverse modeling approaches is capturing changes due to a shock (Psycharis et al., 2014; Kitsos and Bishop, 2018), conceived either qualitatively or quantitatively. However, several more degrees of complexity apply along with the modeling choices to the measurement of resilience. One degree concerns the time configuration of the shock (Martin and Sunley 2014; Kitsos and Bishop, 2018), where the reference to the 2008 economic crisis prevails in the literature (Brakman et al. 2014; Psycharis et al. 2014; Xiao et al., 2017), while some studies go beyond a single reference (Martin et al., 2016; Sensier et al., 2016; Martin and Gardiner, 2018) by examining more than one economic shocks. Another degree of complexity is about the duration of the examined period, where the vast majority of literature studies applies after 1970 (Sensier et al., 2016; Xiao et al., 2017;

Martin and Gardiner, 2018; Kitsos et al., 2019), making economic resilience a matter of the modern economic history. Dealing with these two degrees of complexity is a very important choice for setting a meaningful reference for the conceptualization and measurement of economic resilience (Martin and Sunley, 2014; Kitsos and Bishop, 2018). A third-degree regards the spatial configuration of the economic system (Modica and Reggiani, 2015), which can apply at all levels of geographical scale, international (Brakman et al. 2014; Dijkstra et al. 2015), national (Martin et al., 2016; Martin and Gardiner, 2018), regional (Andreoni and Duriavig, 2013; Sdrolias et al., 2022), or of lower scale (Karrholm, 2014), and drive accordingly to different results. Another degree of complexity deals with the attribute expressed by the dataset, and thus the variable on which resilience is measured. For this choice, a diversity of proxies of structural and production characteristics of the economy is available in the literature, where GDP (Dijkstra et al., 2015; Sensier et al., 2016; Xiao et al., 2017), GVA (Martin and Gardiner, 2018), and employment (Bristow et al., 2014; Martin et al., 2016; Sensier et al., 2016; Kitsos et al., 2019) prevail. The attribute on which resilience is examined is determinative for the outcome of the study and may even drive into opposite results (Sensier et al., 2016).

Within this context of polyphony, the modeling and measurement choices are diverse and at the same time critical in the definition of economic resilience, which promises to contribute to a holistic structural and functional conceptualization in the study of spatial-economic systems. However, this holistic potential does not seem to accordingly apply in the literature, where current synthetic approaches (Schwarz et al., 2011; Palaskas et al., 2015; Sensier et al., 2016; Kitsos and Bishop, 2018; Sdrolias et al., 2022) appear either constrained, mainly due to a pairwise conceptualization, or counterintuitive, due to their high level of complexity and usually apply at geographical scales from international and below. As far as the global scale is concerned, papers dealing with this topic are restricted in number and cover diverse topics, such as economic resilience in the 2008 world trade collapse (van Bergeijk, 2017), concerning economic vulnerability (Zaman and Vasile, 2014; Briguglio, 2017), in the context of bibliometric analysis (Shymon, 2020), or in providing a conceptual (not yet quantitative) context of economic resilience in the era of COVID-19 (Jenny, 2020; Hynes et al., 2020). A common aspect that can be found in these diverse researches is a consensus that the international dimension in the economic resilience empirical research has not been examined in a comprehensive context, necessitating further research. Towards serving this demand, this paper develops a composite (3D) index for measuring economic resilience, conceived in an integrated engineering, ecological, and evolutionary context, and provides an application on GDP temporal data, available for 200 countries of the globe.

By developing the 3D index, this paper aims in contributing to the requirement of a holistic measurement for economic resilience, and by conducting an application on the global scale aims in contributing in incorporate the international dimension in relevant research, as it is described in detail at the following section.

## 3. METHODOLOGY AND DATA

The methodological framework of this paper builds on a multilevel consideration consisting of five steps. In the first step, we develop a novel three-dimensional (3D) index for the measurement of economic resilience, where each dimension conceives one of the diverse conceptual aspects of engineering, ecological, and evolutionary resilience. In the second step, we extract from the literature of modern (after 1970) economic history several crises' milestones to provide references to shocks, for the definition of economic resilience. In the third step, we compute the proposed index about the available shocks and we, therefore, configure a spatiotemporal database for applying further statistical testing and empirical analysis. Also, along with the economic resilience variables, we configure the set of socioeconomic variables participating in the empirical analysis. In the fourth step, we examine the numerical and socioeconomic performance of the proposed index, by testing its ability to provide discrete results amongst its components and detecting correlations between the resilience index and socioeconomic aspects. Finally, in the fifth step, we map the spatial distributions of the proposed resilience index and we detect geographical differentiation in the available aspects of economic resilience captured by each component. Overall, the analysis aims to provide insights into the global geography of resilience to the modern financial and economic crises and its socioeconomic determinants.

### ■ The 3D Resilience Index

In the first step, we develop a three-dimensional (3D) index of economic resilience, building on the context of capturing changes due to a shock (Martin and Sunley 2014; Psycharis et al., 2014; Kitsos and Bishop, 2018). Under the relevant literature (Martin and Sunley, 2014; Kitsos and Bishop, 2018; Sdrolias et al., 2022), a critical choice for defining each component of the proposed vector index is to set a meaningful time reference for the conceptualization and measurement of economic resilience. Towards this direction, we define two periods: one before the shock, called reference period (*R*) and a second one after the shock, performance period (*P*), as it is shown in Fig.1.

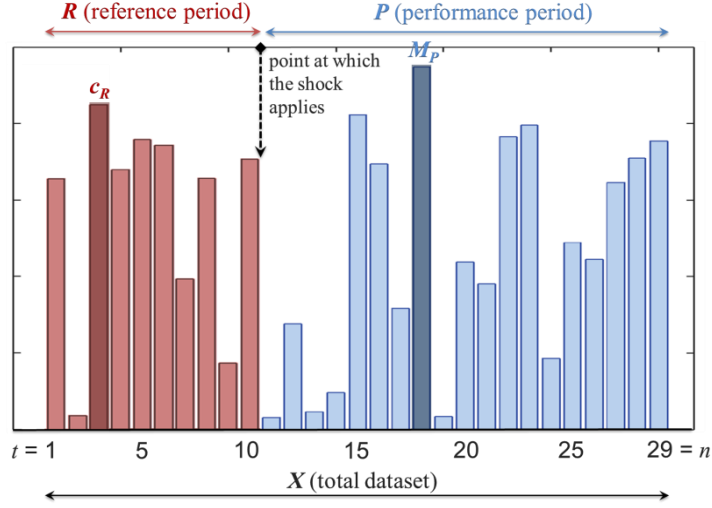

**Fig.1.** A shock divides the time evolution of a proxy attribute $X$ into a reference ($R$) and a performance ($P$) period, where we can define their concordant levels ($c_R$, $M_R$).

Within this context, the proposed index $I_R$ is defined in a vector form by the mathematical formula:

$$I_R = (R_{en}, R_{ec}, R_{ev}) \qquad (1),$$

where:

- $R_{en}$ is the engineering component, which measures the recovery speed of a system due to a shock and is defined by the formula

$$R_{en} = \frac{\log\left(\frac{n - t(c_R)}{t(x_{i \in P} \geq c_R) - t(c_R)}\right)}{\log(n - t(c_R))},$$

with $n$ the length (number of elements) of the dataset $X$; $t(x_i)$ the time point for attribute score $X = x_i$, $R$, $P$ are the reference and performance periods; $c_R$ the level of functionality in the reference period, defined by the maximum $c_R = \max\{X_R\}$; and $M_P$ the level of functionality in the performance period, defined by the maximum $M_P = \max\{X_P\}$

- $R_{ec}$ is the ecological component, which expresses the difference in the level of a system's functionality before and after the shock and is defined by the formula

$$R_{ec} = \left(\exp\left\{\frac{|M_P - c_R|}{\max\{|c_R|, |M_P|\}}\right\}\right)^{\operatorname{sgn}^*\{c_R - M_P\}},$$

with $|M_P - c_R|$ the absolute difference between the levels of functionality before and after the shock; and $\operatorname{sgn}^*\{c_R - M_P\}$ a sign (signum) function, adjusted to yield also a positive sign to zeros ($0 \equiv +1$), and

- $R_{ev}$ is the evolutionary component, which measures the adaptability of a system to continuously respond to changes induced by a shock and is defined by the formula

$$R_{ev} = \exp\left\{-\sum_{i \in P} \frac{c_R - x_i}{n(P) \cdot |c_R + M_P|}\right\}, \text{ with}$$

$n(P)$ the length (number of elements) of the performance period, and $c_R - x_i$ the difference between the level of functionality in the reference period and the level of functionality in every step of the performance period.

For considering an aggregate, one-dimensional, version of the 3D economic resilience index ($I_R$), we introduce a scalar function $I_R : \mathbb{R} \to \mathbb{R}^3$ defined by the Euclidean norm (2-norm) of the indicator components, according to the formula:

$$I_R(\boldsymbol{I_R}, c_R) = \frac{\text{sgn}^*\{c_R - M_P\}}{\sqrt{3}} \cdot \left\| (R_{en}, R_{ec}, R_{ev}) \right\|_2 \qquad (2),$$

where $\left\| (\cdot) \right\|_2$ is the Euclidean 2-norm. A detailed description of the proposed index of economic resilience is available in the Appendix (A.1.).

## ■ Selection of Reference Crises (1970-2020)

In the second step of the methodological framework, we extract from the relevant literature several crises of the modern economic history, to provide references to the shocks defining economic resilience on the conceptual basis shown in Fig.1. In particular, this paper conceives the last half-century as the time of modern economic history, and therefore the collected crises' milestones refer to the period 1970-2020. Provided that the setting of a meaningful reference as a shock is important for the definition and measurement of economic resilience (Martin and Sunley, 2014; Kitsos and Bishop, 2018), this paper goes beyond current approaches examining mainly one (up to four) economic crisis (see literature review), and takes under consideration as much as possible milestones of the economic decline occurred in the last half-century. In this paper, the criteria for selecting the economic crises included in the analysis are, first, the geographical scale of their economic impact (to the extent it can be approximated and estimated from the available sources) and, secondly, their availability in scientific literature. Such pluralism in the configuration of shocks first provides plenty of degrees of freedom in computations and thus is expected to facilitate the exploratory and empirical evaluation of the 3D economic resilience index's performance. Next, this abundance of shocks counterbalances the requirement of thoroughly reviewing and evaluating the importance of economic crises in the context of political economy and economic history, as would be required in the case of selecting just a few crises according to their importance. Besides, evaluating the importance of economic crises in terms of political economy and economic history falls out of the scope of this paper. Within this context, fifteen (15) milestones of financial and economic crises are extracted from the literature (although their number may not be exhaustive for the reference period), as is shown in Table 1, where a short description and estimation of the geographical scale of the crises economic impact is included.

**Table 1**

List of major modern financial crises (1970-2020)*

| Reference Year | Crisis | Reference |
|---|---|---|
| 1973 | Oil crisis | Akins (1973); Issawi (1978); Mitchell (2010) |
| 1980-1982 | Latin American debt crisis | Wiesner (1985); Sims and Romero (2013) |
| 1987 | Black Monday | Markham and Stephanz (1987); Schaede (1991); Metz (2003) |
| 1988 | Norwegian banking crisis | Englund (1999); Ongena et al. (2003) |
| 1989 | United States Savings & Loan crisis | Pyle (1995); Curry and Shibut (2000); Barth et al. (2006) |
| 1990 | Japanese financial crisis (asset price bubble) | Hoshi and Kashyap (2004) |
| 1991 | Scandinavian banking crisis | Englund (1999); Ongena et al. (2003) |
| 1992-1993 | Black Wednesday | Fratianni and Artis (1996); Soderlin (2000) |
| 1994 | The economic crisis in Mexico | Carstens and Schwartz (1998); Davis-Friday and Gordon (2005) |
| 1997 | Asian Financial Crisis | Maroney et al. (2004): Deesomsak et al. (2009) |
| 1998 | Russian financial crisis | Buchs (1999); Feridun (2004) |
| 1999 | Argentina economic crisis | Bebczuk and Galindo (2008) |
| 2000 | Turkish economic crisis | Alper (2001); Cizre and Yeldan, (2005) |
| 2001 | Bursting of the dot-com bubble | Ljungqvist and Wilhelm (2003); Goodnight and Green (2010) |
| 2007-2008 | Worldwide financial crisis | French et al. (2009); Luchtenberg and Vu (2015) |

*. For a detailed description, see Appendix (Table A1)

■ **Data and Variable Configuration**

In the third step of the methodological framework, we compute the proposed 3D index about the available shocks (on the rationale shown in Table 1). The 3D index $I_R=(R_{en}, R_{ec}, R_{ev})$ is computed on GDP data of the period 1960-2020, which are available for 200 countries worldwide, measured in US dollars (US$), and were extracted from the World Bank database (Worldbank, 2021). The US$ measurements refer to constant prices of the year 2010. Taking into account the debate about the diversity of proxies, which are available in the literature for the measurement of resilience (Martin et al., 2016; Sensier et al., 2016; Xiao et al., 2017; Kitsos et al., 2019) and highlight the dependency between resilience and its defining attribute (Sensier et al., 2016), this paper defines resilience in terms of GDP due to its popularity in literature, its direct linkage with production, and its availability for more countries. Within this context, the computations of the proposed 3D index $I_R=(R_{en}, R_{ec}, R_{ev})$ of economic resilience yield four different vector variables of length 2800 (=200×14, the number of 200 available countries included in the analysis times the 14 examined crises), one referring to the 1D index of economic resilience $I_R$, and the other three to the engineering, ecological, and evolutionary components of the 3D index. Further, to examine the socioeconomic performance of the proposed index, we include in the analysis a set of fourteen socioeconomic variables shown in Table 2. These variables are selected to capture the main socio-economic attributes that have been found in the relevant literature to correlate with economic resilience, such as sectorial structure and competitiveness (Psycharis et al., 2014; Palaskas et al., 2015; Martin et al., 2016), socioeconomic and demographic configuration (Kitsos et al., 2019), regional urbanization and specialization (Brakman et al., 2014), production quality (Capello et al., 2015), and human capital (Kitsos and Bishop, 2018), and their choice depends on data availability. All variables participating in the analysis have a length of 2800, were extracted from the Worldbank (2021) database, and are organized in panel data consisting of entries of the period 1960-2020 for the 200 available countries.

**Table 2**

List of socioeconomic variables* participating in the analysis

| Name | Description |
|---|---|
| POP | Total population |
| URB | Urban Population |
| URB2 | Population in the largest city |
| LF | Labor force (≥15 years) |
| EMP | Employment to population ratio (≥15 years) |
| GDPpc | GDP per capita (constant 2010 U.S. dollars) |
| GVA | GVA at basic prices (constant 2010 U.S. dollars) |
| AVGA | Prime sector GVA (constant 2010 U.S. dollars) |
| BGVA | Industry GVA (constant 2010 U.S. dollars) |
| CGVA | Services GVA (constant 2010 U.S. dollars) |
| TRD | Trade (% of GDP) |
| FCE | Final consumption expenditure (constant 2010 U.S. dollars) |
| TXR | Tax revenue (% of GDP) |
| TNR | Total natural resources rents (% of GDP) |

*. For a detailed description, see Appendix (Table A2)
Source: Worldbank (2021)

### ■ Numerical and socioeconomic performance of the proposed 3D index

In the fourth step of the methodological framework, we test the numerical and socioeconomic performance of the proposed index, using statistical inference and empirical methods. This part of the analysis statistically tests average numeric equality between a variable's groups, by using confidence interval (CIs) error bars for the mean values. The construction of error bars builds on the rationale of the independent-samples *t*-test (Norusis, 2011; Walpole et al., 2012), used to compare the average values $\mu_\alpha$ and $\mu_\beta$ between a pair of groups configured from the same variable $X$ under an arithmetic (cutting point) or categorical grouping criterion. This test specializes according to the equality of variances between groups, and produces separate results for unpooled and pooled variances according to their significance, based on *Levene's* test (Norusis, 2011). For this paper, instead of applying *t*-tests, we construct error bars of confidence intervals because they are more intuitive, due to their graphical format. In terms of interpretation, non-overlaying error bars of (1–*a*)% CIs imply (1–*a*)% certainty that the detected inequality is not a matter of chance. Further, a concordant to *Levene's* test information about the equality of variance can be extracted from the length of error bars. In the calculations, missing values are excluded pair-wisely (Norusis, 2011), where the algorithm keeps from test-to-test the max possible degrees of freedom (Walpole et al., 2012) to perform the analysis. Finally, to perform this analysis, we group the available variables in accordance to their $I_R$ performance, where we consider: (*i*) the first group consisting of the negative values of the 1D economic resilience index ($I_R<0$), describing a negative ecological performance with simultaneous lack of engineering and perhaps (to be tested) deficient evolutionary performance; (*ii*) the second group consisting of positive values smaller than one (0<$R_{en}$<1), describing a satisfactory overall resilience profile with potential deficiencies to be tested; and (*iii*) the third group consisting of values greater than one ($I_R$>1), illustrating an overall high resilience profile of high recovery speed, with simultaneous positive ecological and relatively high evolutionary performance. Within this context, in the part of the numerical examination, we statistically evaluate the numerical performance of the $R_{en}$, $R_{ec}$, and $R_{ev}$ components in terms of this 3-class $I_R$ grouping, to get insights into the structural differences and configuration of the components. In the final part of this analysis, we empirically examine differences amongst the $I_R$ classes for the available socioeconomic variables shown in Table 2. Statistical differences captured amongst the $I_R$ classes for a socioeconomic variable $X$ imply the existence of distinguishable levels in the attribute $X$ due to the economic resilience's classification, illustrating a statistical relevance between variable $X$ and economic resilience (as captured by the proposed index $I_R$). This approach aims to detect new or evaluate existing determinants of economic resilience that are available in the literature.

■ **Geographical economic evaluation of the proposed 3D index**

In the final step of the methodological framework, we examine the performance of the proposed resilience index in a geographical economic context. To do so, we build on descriptive, statistical inference, and empirical methods to extract this time geographical information from the available data. In particular, examine the spatial distributions of the proposed resilience index and its components in the world map, and we further test any geographical differentiation in the available economic resilience variables based on descriptive (population pyramid) and statistical inference (CI error bars) techniques. The overall approach has a double purpose; first to further evaluate the performance of the proposed index, and secondly to get further insights into the relationship between geography and resilience, which is known to lie within a symbiotic context.

## 4. RESULTS AND DISCUSSION
### 4.1. Examining the numerical and socioeconomic performance of the proposed 3D index

In the first part of the analysis, we test the numerical and socioeconomic performance of the proposed 3D index, and the results are shown in Fig.2 (numerical evaluation) and Fig.3 (socioeconomic evaluation). As it can be observed in Fig.2, for both confidence levels (95% and 99%), the CIs of each component amongst the $I_R$ classes lie in a discrete range, following an ordering of the form $R_i(I_R<0)<R_i(0<I_R<1)<R_i(I_R≥1)$, for $i=en, ec, ev$. This ordering is concordant to the inequality $\{I_R<0\}<\{0<I_R<1\}<\{I_R≥1\}$, illustrating a monotonous performance of the $R_{en}$, $R_{ec}$, and $R_{ec}$ components under the monotonicity of the 1D index.

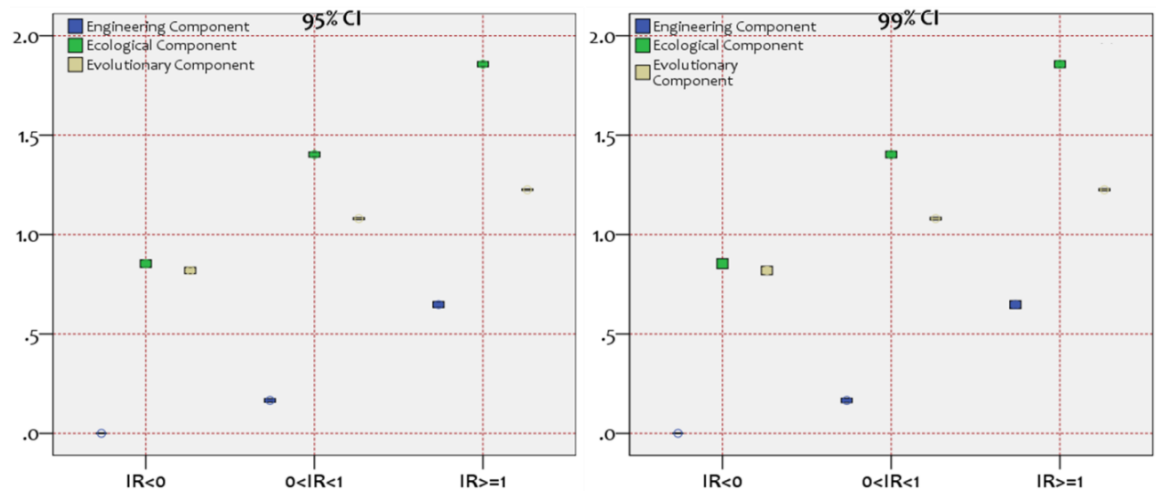

**Fig.2.** Error bars of 95% CIs (left) and 99% CIs (left), for the mean values of the $I_R$'s components, across the classification $\{I_R<0\},\{0<I_R<1\},\{I_R≥1\}$ of the proposed 1D index.

On the other hand, inequalities within the $I_R$ groups follow a different, but also standard, ordering of the form $R_{en}(j)<R_{ev}(j)<R_{ec}(j)$, for $j=\{I_R<0\},\{0<I_R<1\}$, and $\{I_R≥1\}$, producing distinguishable CIs in all cases except the inequality $R_{ev}(I_R<0)<R_{ec}(I_R<0)$, where the concordant CIs slightly overlay (although not for 90% CIs). This result illustrates discrete magnitudes of numerical scale for each component within each $I_R$ class, also expressing a good attribute in the performance of the proposed index.

In the part of the socioeconomic evaluation, Fig.3 shows differences amongst mean values of the available socioeconomic variables (Table 2) across the classification $\{I_R<0\},\{0<I_R<1\},\{I_R\geq1\}$ of the proposed 1D index. First, Fig.3i shows that countries with a high overall economic resilience ($I_R\geq1$) also have a high population. This significant outcome is in line with classic theories on economic growth, gravitational economies, and economies of scale (Behrens and Thisse, 2007; Polyzos, 2019), as long as with specialized works (Hallegatte, 2014) highlighting the positive effect of the population in the configuration of economic resilience.

Next, Fig.3ii shows that countries with a high overall economic resilience have also a significantly high urban population. However, since Fig.3iii does not reveal any significant outcome for the population in the largest city, a combined consideration with the previous finding reveals that economic resilience is more related to peripheral than centralized urban structures (where most of the population is concentrated in the largest city). This outcome supports the finding of Brakman et al. (2014), who observed that commuting areas (those intermediating cities and rural areas) are relatively more resilient than cities and rural areas, per se. Next, Fig.3iv illustrates a tendency of countries with high economic resilience to have an also a high labor force, while (impressively) countries of intermediate (or just good) resilience are described by a low labor force.

Next, Fig.3v describes that countries with a high economic resilience significantly have an also high share of employment. Although detecting a relevance between economic resilience and employment is somewhat expected, these findings reveal potential linkages between production-based and employment-based measurement approaches (Martin et al., 2016; Kitsos et al., 2019) of economic resilience. Next, figures Fig.3vi-xii evaluate the performance of the proposed 1D index in terms of productivity and specialization. For the cases of GDP (Fig.3vii) and GVA (Fig.3viii), we can observe similarities indicating significant differences between the high resilience ($I_R\geq1$) and negative resilience ($I_R<0$) groups. We can also observe a similarity in the pattern between these two variables and the share of employment (Fig.3v), further supporting a previous comment on the linkages between productivity and employment in the measurement of economic resilience.

Next, amongst the aspects of production specialization (Fig.3ix-Fig.3xii), only the cases of agriculture (Fig.3ix) and trade (Fig.3xii) provide significant results. On the one hand, Fig.3ix illustrates that countries with a high overall economic resilience have also a high specialization in agricultural production. This significant result can be striking to the extent that relates economic resilience in the agricultural sector with the inelastic demand of agricultural goods (Sdrolias et al., 2022), which serve basic needs, although the sector is sensitive to environmental externalities, such as climate conditions. On the other hand, Fig.3xii shows that countries with a high overall economic resilience have a relatively low share of trade. This significant result is striking whether observing that the CI of the low resilience group ($I_R$<0) is above 100%, illustrating an importing sign in the trade balance. This interpretation implies that economically resilient countries are likely to be self-sufficient, while non-resilient countries are not. Next, the error bar patterns of consumption (Fig.3xiii) and tax revenue (Fig.3xiv) are similar to the GDP's (Fig.3vii), illustrating a positive incline of these variables as economic resilience grows. In terms of consumption, this outcome is due to the GDP definition (Polyzos, 2019) including (among others) a consumption term. As far as tax revenue is concerned, this result is due to the proportional income-based definition of tax revenue (Polyzos, 2019). Finally, Fig.3xv illustrates that countries with a low overall economic resilience have also a significantly low share in natural resources. This outcome highlights the positive aspect of sustainability in the context of economic resilience (Rose, 2007), but the effect of natural resources (let it denoted with $X$) in this relationship does not appear linear, according to the inequality

$$\mu_X(I_R<0)<\mu_X(I_R\geq1)\leq\mu_X(0<I_R<1),$$

where $\mu$=average.

This non-linear pattern implies that the human added value in production appears more important in the configuration of economic resilience compared to the natural resources, although the latter is also of high importance. Finally, in terms of heterogeneity, the analysis in Fig.3, in all cases except population (Fig.3i) and natural resources (Fig.3xv), shows that the first class ($I_R$<0) enjoys the longest CIs. To the extent that it does not depend on bias due to the sample size (where the 2800 available cases of panel data are unevenly distributed across the classes, with odds 1:14:40), this observation indicates that the low resilience groups ($I_R$<0) are described by the highest heterogeneity. In conceptual terms, this interpretation can also be seen as a high diversity in the determinants configuring cases of low economic resilience, addressing avenues of further research towards this direction.

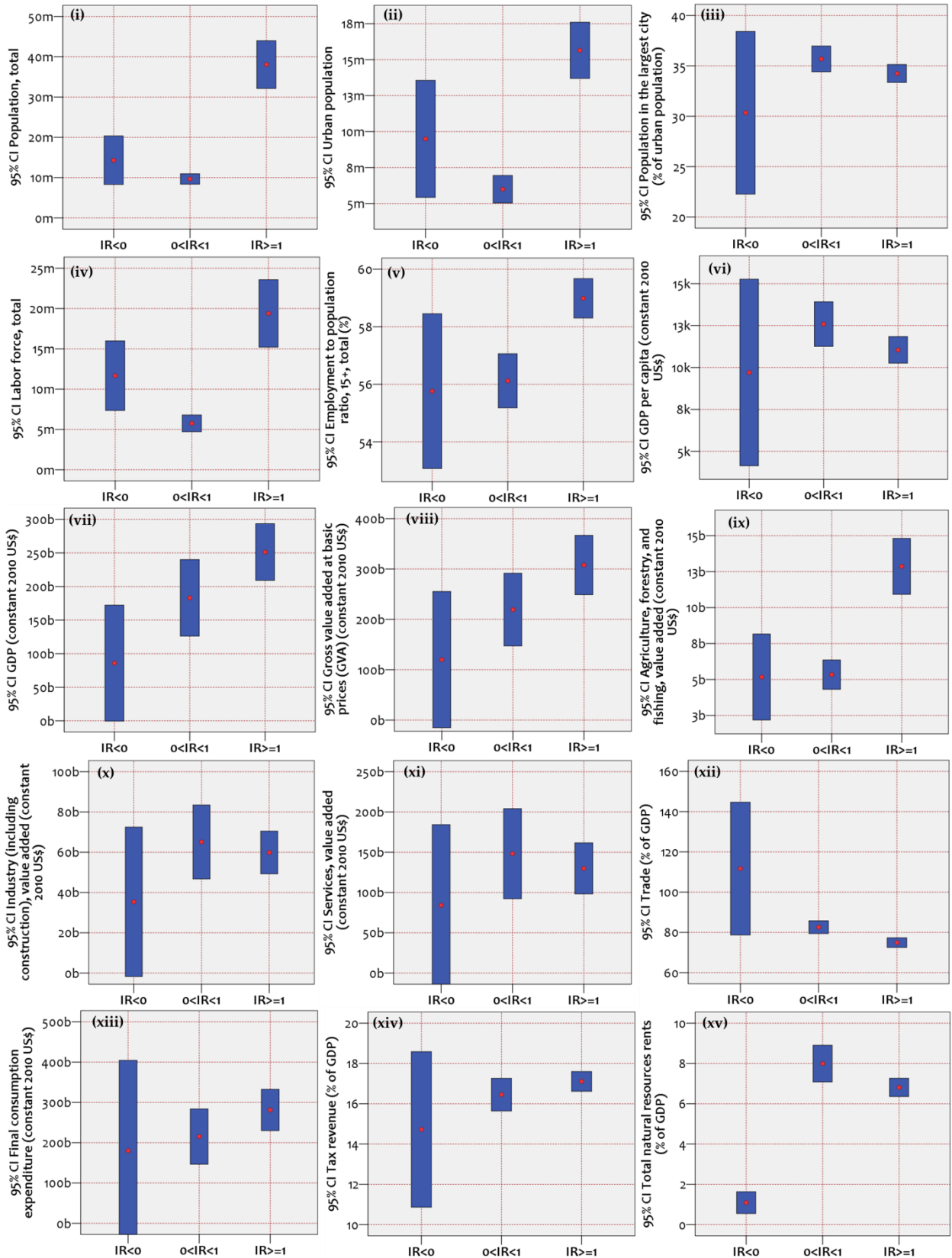

**Fig.3.** Error bars of 95% CIs for the mean values of the socioeconomic variables of Table 2, across the classification $\{I_R<0\},\{0<I_R<1\},\{I_R\geq 1\}$ of the proposed 1D index.

## 4.2. Examining the geographical economic performance of the proposed 3D index

In the second part of the analysis, we examine the performance of the proposed resilience index within the context of regional science and applied geography. At a first approach, we extract from the available panel data the countries that remained in the same class (fixed classes) during available shocks and those that commuted (commute classes) by either demoting or promoting from one to another class. Countries with fixed classes are 37.5% of the total, whereas those commuted are 59% of the total. The geographical distribution of the fixed classes is shown in the map of Fig.4, where it can first be observed that the first class ($I_R<0$) has no fixed cases. This observation can provide a positive aspect in the evolution of economic resilience through time. As far as the other two classes are concerned, we can observe that their spatial distribution on the map shapes two distinguishable horizontal patterns, configured about the latitude crossing China. In particular, countries of high resilience ($I_R≥1$) are distributed southern the latitude crossing China, while countries of the second class ($0<I_R<1$) are distributed at northern latitudes than China. Interesting exceptions in this pattern of geographical duality are (*i*) in Africa, the cases of Liberia (LBR) and Libyan Arab Jamahiriya (LBY); and (*ii*) in Asia, the cases of Israel (ISR), Tajikistan (TJK), and Kyrgyzstan (KGZ), belonging in the medium instead of the high class of economic resilience. Although these countries as very diverse in terms of geomorphology and economic profiles, political instability (Fielding, 2003; Moran and Buell, 2013; Sulaimanova and Bostan, 2014; Al-Shamamri and Willoughby, 2019) can be seen as a common feature in these cases, its correlation with economic resilience does not lie within the scope of this paper, addressing avenues of further research. In terms of geographical clustering, we can observe that cases of high-class form clusters are continental defined, except Asia and Oceania which can be seen as one cluster. For the case of the medium class, we can observe clearly defined continent clusters, located in northern America, east Europe, and northern Asia, along with the previously mentioned exemptions.

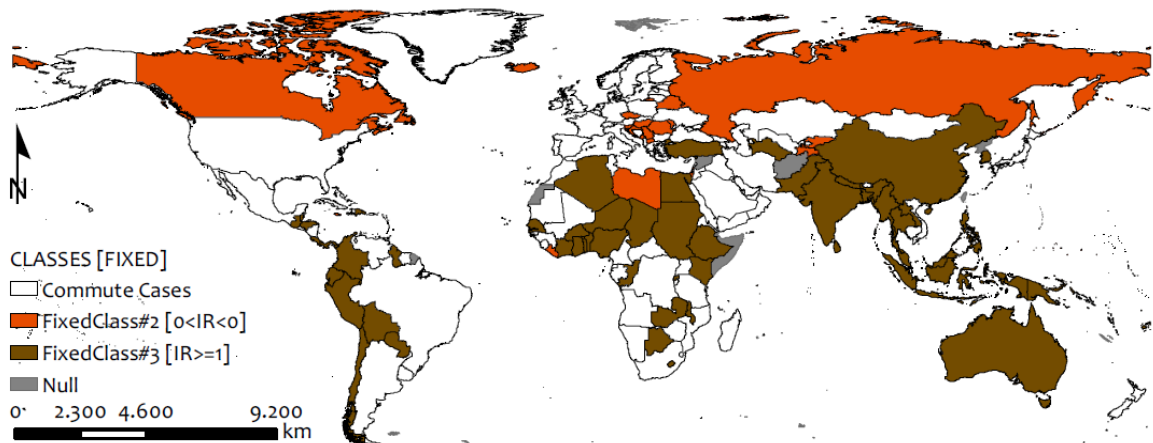

**Fig.4.** Countries preserved their economic resilience class during the period 1960-2020.

Next, the geographical distribution of the commute classes is shown in Fig.5. Although this map is configured by cases complementary to the fixed classes, the spatial patterns emerging here are far from shaping the previous discrete horizontal zoning. In particular, we can observe as cases of average low economic resilience the intercontinental dipole Georgia (GEO) - Ukraine (UKR), located in eastern Europe and West Asia, and the country of Kiribati (KIR) in Oceania. Although they were commute cases in the examined period, these countries belong on average to the first class ($I_R<0$) of low resilience. Amongst the diverse features describing the geomorphological and socioeconomic profile of these countries, a distinguishable common feature can also be found in the context of political history and economy, regarding the recent independence of Kiribati (in 1971), Georgia (in 1991), and Ukraine (in 1991), and the political instability describing their modern history (Storey and Hunter, 2010; Nilsson and Silander, 2016). Next, the spatial distributions of the other two classes of economic resilience deviate from a clearly defined linearity, as the case of the fixed classes is. Perhaps the only distinguishable linear pattern appears in southwest Asia and northeast Africa, shaping a linear arrangement of alternating 2nd class and 3rd class cases. In all the other cases, we can observe centrifugal patterns consisting of a 2nd class core and a 3rd class periphery, such as in America, South Africa, and Europe.

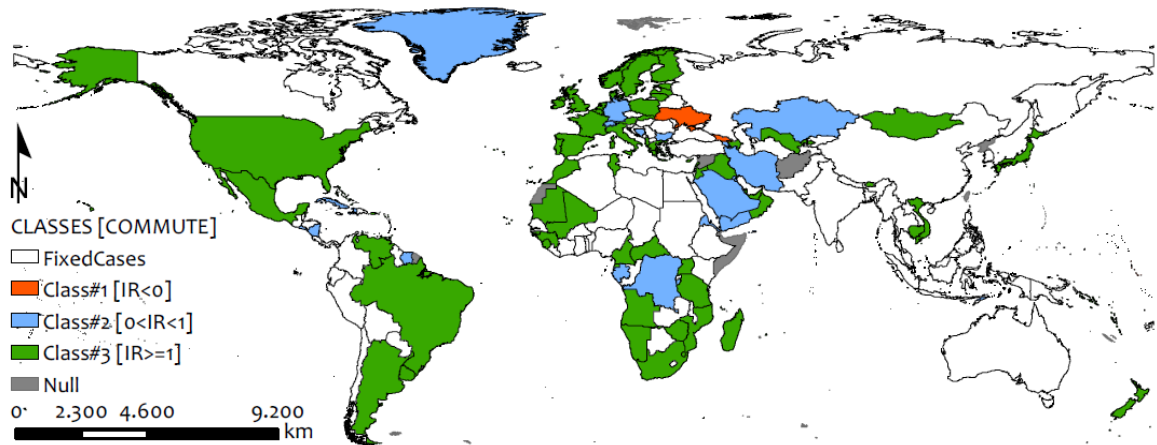

**Fig.5.** Countries commuted between economic resilience classes during the period 1960-2020 (classes computed by averages, rounded to integers)

To complement the previous analysis, we construct the population pyramid of Fig.6, showing the sum of shifts (adding –1 for demoted and +1 for promoted cases) in the classes of economic resilience along the time sequence of the available shocks and per continent. Due to the definition of shifts as differences, one degree of freedom (for the 1973 shock) is lost. Provided that the effect of each shock is traceable in the year of its emergence, this approach allows evaluating indirectly the effect of each shock on each continent. In particular, we can observe that the economic resilience of Africa was not affected by the 1989 (United States savings & loan), 1990 (Japanese), and 2000 (Turkish) crises, while it appeared to benefit from 1994 (Mexico) and 1997 (Asian) crises by recording positive shifts. Next, the economic resilience of Asia was not affected by any shock during the period 1987-1994 (shocks external to the continent except 1990 in Japan) and by 1999 (Argentina) crisis, while it appeared to benefit from the 2000 (Turkish) crisis. As far as the economic resilience of Europe is concerned, it was not affected by the 1988 (Norwegian) and 1990 (Japanese) crises, by the crises of the period 1992-1994, and by 1999 (Argentina) crisis, while it appeared to benefit from the 1991 (Scandinavian) crisis. Next, the economic resilience of North America was not affected by the 1982 (Latin American) crisis, by the crises of the period 1991-1993 (shocks external to the continent), as well as by 1999 (Argentina) and 2000 (Turkish) crises, while it appeared to benefit from the 1997 (Asian) crisis. Next, the economic resilience of Oceania was not affected by the 1982 (Latin American) crisis, by the crises of the period 1988-1991 (shocks external to the continent), and by the crises of the period 1998-2008 (shocks external to the continent), but also it did not benefit by any case. Finally, the economic resilience of South America shows an impressively steady performance since it recorded demotions only in the shocks of 1980 (endogenous Latin American crisis), 1998 (Russian financial crisis resulted in the next year in Argentina's crisis), and in 2008 (worldwide crisis), while it appeared to benefit from the 2001 (dot-com bubble) crisis. As far as the 2008 economic crisis is concerned, Europe seems to be affected the most, followed by Africa, Asia, North America, and South America, while Oceania appears indifferent to the effect of the recent global economic crisis. Overall, in the majority of cases (Africa, North and South America, and Oceania), the good temporal performance in economic resilience classification was due to the geographical distance from shocks' origin, while Asia and Europe were also benefited from internal crises.

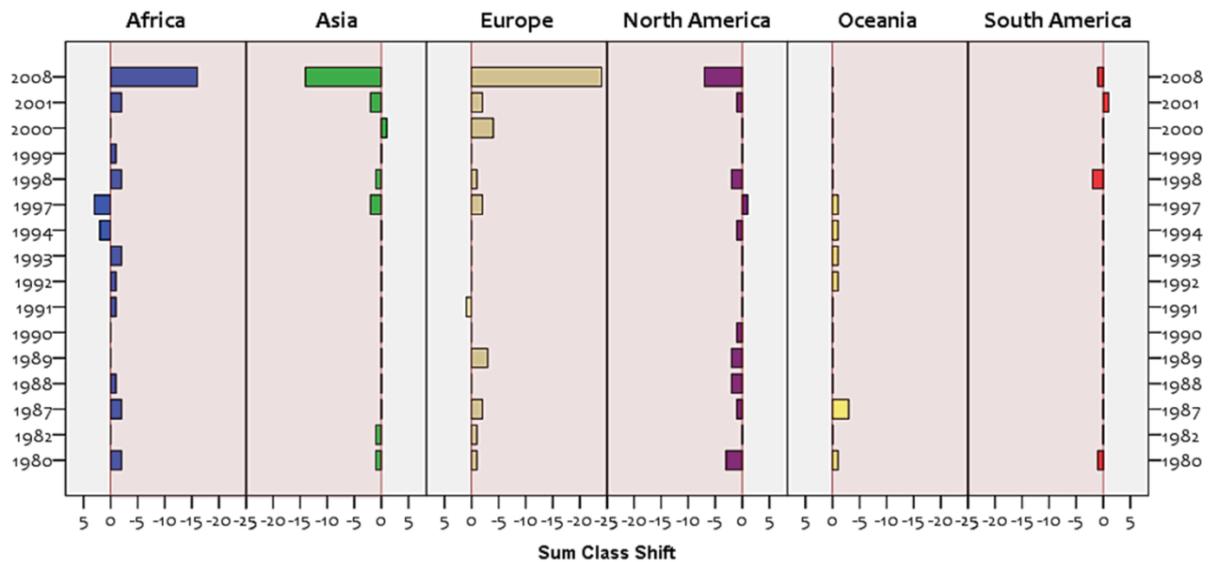

**Fig.6.** Population pyramid showing the sum of shifts in the classes of economic resilience, distributed along the available reference years.

At the final approach, we detect statistical differences in the mean values amongst continent groups of the proposed 1D index of economic resilience and its engineering, ecological, and evolutionary components, based on error bars computed on the total data length (2800). The results of the analysis are shown Fig.7, where we can observe three patterns of inequality: one concerning the engineering component, described by the descending order South America, Africa, North America, Asia, Europe, and Oceania; a second one concerning the other two components, described by the descending order Asia, Africa, South America, North America, Europe, and Oceania; and a third one describing the 1D index performance, averaging the previous two, described by the descending order Asia, South America, Africa, North America, Europe, and Oceania. In all three patterns, Europe, and Oceania are the last. This result is in line with the previous findings of Fig.5, according to which these continents are the only with cases of commute low class ($I_R<0$).

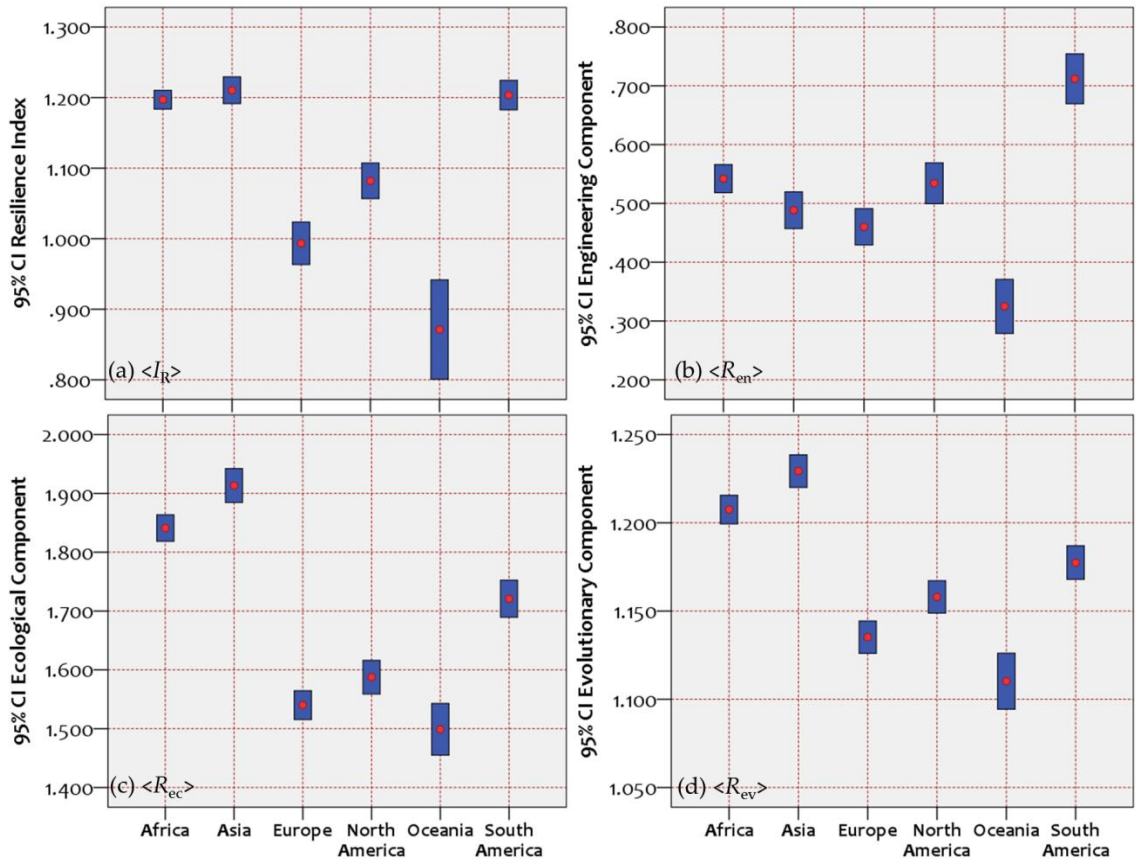

**Fig.7.** Error bars of 95% CIs for the mean values of the proposed 1D index of economic resilience and its engineering, ecological, and evolutionary components, per continent groups.

As far as the other continents are concerned, their order is shuffled in these three inequality patterns, implying a diverse performance specialized in the attribute expressed by each component. For instance, South America has the significantly highest average engineering resilience, implying the best performance in recovery speed from shocks, a fact which complies with the findings of Fig.6 showing that this continent was affected the least by the examined shocks. Similarly, North America has relatively a high average engineering resilience compared to its ecological and engineering performance. This outcome is also in line with the findings of Fig.6 showing a satisfactorily low effect of the examined shocks. On the other hand, Africa and Asia are cases of significantly highest ecological and evolutionary resilience, illustrating on average that these continents are highly adaptable and capable of shifting to states of higher functionality due to the shocks, a fact that was supported by the analysis in Fig.4 and Fig.5.

## 4.3. Discussion

Within the high complexity describing the conceptual and measurement context of economic resilience, this paper introduced a novel 3D index integrating the diverse equilibrium (engineering and ecological) and evolutionary aspects of economic resilience in a single measure. The proposed 3D index includes a component for each aspect (engineering, ecological, evolutionary) of economic resilience and was examined and evaluated for its numerical, socioeconomic, and geographical economic performance. The numerical analysis reveals the monotonicity of all components under the monotonicity of the 1D index (computed by averaging the 3D index) and also revealed the good performance of producing discrete magnitudes of numerical scale for each component. Next, the socioeconomic analysis provides insights into correlating the classes of low (#1: $I_R<0$), medium to good (#2: $0<I_R<1$), and high (#3: $1<I_R$) economic resilience with various socioeconomic macroeconomic characteristics each and allowed to collectively configure the socioeconomic profile of each class. As it can be observed in Table 3, the first class of low economic resilience includes countries of low population and urban population, low agricultural specialization and share of natural resources, and a high importing profile of trade balance. In this class, no cases remained fixed through time. However, countries that on average commuted in the low resilience class during the examined period are just three countries, exclusively originating from Europe (GEO, UKR) and Oceania (KIR).

Next, the second class of moderate to good economic resilience includes countries of low population and urban population, low labor force, and agricultural specialization, but with a high share of natural resources and a high importing profile in the trade balance. In comparison with the previous class, the second class just underperforms in the labor force and outperforms in natural resources. In terms of production, this observation implies that the labor force is not a major determinant in the setting of a medium to good economic resilience, but the capacity of natural resources is. In the second class of economic resilience, 11% of the total countries (sub-table B) remained fixed, while 15.5% commuted in this class during the examined period. In terms of geography, the class includes 26.5% of the total countries, originating over 75% from Europe, Asia, and North America.

### Table 3
Tabulation of the main results of the analysis

| Attribute | Class#1 ($I_R<0$) | Class#2 ($0<I_R<1$) | Class#3 ($I_R\geq1$) | Class#1 ($I_R<0$) | Class#2 ($0<I_R<1$) | Class#3 ($I_R\geq1$) |
|---|---|---|---|---|---|---|
| A. Socioeconomic Analysis (tabulation) ||||||| 
| Population | 🟥 | 🟥 | 🟩 | | | |
| Urban Population | 🟥 | 🟥 | 🟩 | | LEGEND | |

| Attribute | Class#1 (I_R<0) | Class#2 (0<I_R<1) | Class#3 (I_R≥1) | Class#1 (I_R<0) | Class#2 (0<I_R<1) | Class#3 (I_R≥1) |
|---|---|---|---|---|---|---|
| Population in the largest city | | | | 🟩 | Sig. highest scores* | |
| Labor force | | 🟥 | | 🟥 | Sig. lowest scores* | |
| Employment to population ratio | | | 🟩 | ⬜ | Insignificant scores | |
| GDP per capita | | | | 🟨 | Sig. highest scores** | |
| GVA at basic prices | | | | *. 5%; **. 10% level of significance | | |
| Prime sector GVA | 🟥 | 🟥 | 🟩 | >85% | Max | >21% |
| Industry GVA | | | | 70-85% | ↑ | 17-21% |
| Services GVA | | | | 55-70% | | 14-17% |
| Trade (% of GDP) | 🟩 | 🟩 | 🟥 | 40-55% | | 10-14% |
| Final consumption expenditure | | | | 25-40% | | 7-10% |
| Tax revenue (% of GDP) | | | | 10-25% | ↓ | 3-7% |
| Total natural resources rents (% of GDP) | 🟥 | 🟩 | 🟩 | 0-10% | Min | 0-3% |

| B. Geographical Economic Analysis (cross-tabulation) | | | | | | |
|---|---|---|---|---|---|---|
| Fixed Performance | Within continents distrib. (→) | | | Within classes distribution (↓) | | |
| Global | – | 11.0%* | 26.5%* | – | 29.3%** | 70.7%** |
| Africa | | 8.7% | 91.3% | | 2.7% | 28.0% |
| Asia | | 21.1% | 79.0% | | 5.3% | 20.0% |
| Europe | | 92.3% | 7.7% | | 16.0% | 1.3% |
| North America | | 25.0% | 75.0% | | 2.7% | 8.0% |
| Oceania | | 40.0% | 60.0% | | 2.7% | 4.0% |
| South America | | | 100.0% | | | 9.3% |
| Commute Performance | Within continents distrib. (→) | | | Within classes distribution (↓) | | |
| Global | 1.5%* | 15.5%* | 42.0%* | 2.5%** | 26.3%** | 71.2%** |
| Africa | | 14.3% | 85.7% | | 3.4% | 20.3% |
| Asia | | 27.3% | 72.7% | | 5.1% | 13.6% |
| Europe | 5.6% | 19.4% | 75.0% | 1.7% | 5.9% | 22.9% |
| North America | | 47.4% | 52.6% | | 7.6% | 8.5% |
| Oceania | 12.5% | 50.0% | 37.5% | 0.8% | 3.4% | 2.5% |
| South America | | 20.0% | 80.0% | | 0.8% | 3.4% |

*. Computations on the total cases
**. Computations within each category (fixed, commute)

Finally, the third class of high economic resilience includes countries of high population and urban population distributed more peripherally than concentrated in the highest populated city, high levels of employment and agricultural specialization, a high share of natural resources, and a good profile (although the lowest) in the trade balance. Compared to the other two classes, this class has the highest overall significant scores and additionally the highest share of employment, except the GDP share in the trade that is the lowest. As far as employment is concerned, this result underlines the importance of employment in the setting of high economic resilience, as is evident in the relevant literature (Bristow et al., 2014; Martin et al., 2016; Sensier et al., 2016; Kitsos et al., 2019). In terms of GDP share in trade, the score of the third class is the only one below 100%, illustrating a self-sufficient rather than a deficient profile in the trade balance. In the third class, 26.5% of the total countries (sub-table B) remained fixed, while 42% commuted in this class during the examined period. In terms of geography, the third class includes 68.5% of the total countries, originating in amount 75% from Asia, Africa, and Europe.

Overall, the socioeconomic analysis shows that, although is defined in a complex structural and spatiotemporal context, economic resilience is positively related to primary determinants of economic development, such as the gravitational variables of population and decentralized urban population, the productivity coefficients of employment and natural resources, and the inelastic sectorial setting of agricultural specialization. These findings are in line with current literature that highlights (*i*) the positive effect of the population in the configuration of economic resilience (Hallegatte, 2014), (*ii*) the importance of areas intermediating cities and rural areas in shaping the performance of good resilience (Brakman et al., 2014), (*iii*) the determinative role of employment in the definition of economic resilience (Martin et al., 2016; Kitsos et al., 2019), and (*iv*) the positive effect of agriculture in the ability to withstand and recover from shocks (Giannakis and Bruggeman, 2020). Moreover, the socioeconomic analysis shows that economic resilience is related to a self-sufficient profile in the trade balance, going beyond current literature findings (Wand and Wei, 2021) highlighting the role of trade openness and trade dependence in the improvement of economic resilience, and detecting a relationship between low performance in economic resilience and structural heterogeneity, supporting relevant findings of Giannakis and Bruggeman (2020) at a lower geographical scale of European Union.

In terms of geography, this paper fills the literature gap by shaping the worldwide map of economic resilience, detecting a geographical duality between countries of stably high and stably medium to good economic resilience. This duality is defined about the latitude crossing China, revealing that a stably good performance of high economic resilience is continentally defined and is more a matter of the northern than the southern hemisphere. Further, the analysis detects centrifugal patterns of commute performance of economic resilience located in America, South Africa, and Europe. These patterns can be loosely seen as markets of unsteady economic interaction, addressing avenues of further research on the determinants inducing such instability. In the time length of all available crises, this paper also promotes academic dialogue by revealing that a good diachronically performance in economic resilience is mainly due to the geographical distance from the shocks' origin, however, Asia and Europe were also benefited in some cases by internal crises. In the long run, Europe and Oceania appeared the last in their economic resilience performance. As far as Europe is concerned, this result adds in current literature findings (Rivera, 2012; Holl, 2017; Cainelli et al., 2018) underlying the EU's vulnerability due to specialization of manufacturing and construction, along with the diversity in economic resilience between the euro area and non-euro regions. America appeared with relatively better performance in engineering resilience, implying good ability in recovering from shocks, while Africa and Asia with high ecological and evolutionary resilience, illustrating high adaptability and capability to shift to states of higher functionality due to shocks. The analysis also shows that Europe was most affected by the 2008 crisis, Africa, Asia, North America, South America, and Oceania, accrediting current literature in economic resilience that mainly focuses on the 2008 crisis in European Union.

Finally, in an attempt to detect a common feature amongst countries diachronically included in the lowest class of economic resilience, this paper observes political instability as the main determinant of low economic resilience. However, this observation introduces more a research hypothesis than a finding, addressing avenues of further research into the determinants of economic resilience. More avenues of further research can suggest applications of the proposed 3D index on more variables (employment, GVA, etc.) or on data of higher geographical resolution, the construction of models in economic resilience incorporating variables of the proposed 3D index, the development of decomposition techniques, and perhaps the evaluation of the proposed index as a metric of time-series analysis.

## 5. CONCLUSIONS

Towards serving the demand of integration in the measurement of economic resilience, this paper developed a three-dimensional (3D) index, consisting of components capturing diverse engineering, ecological, and evolutionary aspects of economic resilience. The proposed index was computed on GDP data of 200 countries worldwide, for the period 1960-2020, concerning 14 crises considered as shocks, and was found monotonous and of distinguishable numerical scales amongst its components. In addition to its methodological added value, the application of the proposed index on real-world data allowed considering a novel classification of countries about economic resilience, provided insights into the drivers affecting this property, and revealed geographical patterns of economic resilience at the global scale. In contrast to the complexity describing its conceptualization and measurement, economic resilience was found positively related to primary determinants of economic development, such as the gravitational variables of population and decentralized urban population, the productivity coefficients of employment and natural resources, and the agricultural specialization related to the service of inelastic needs. On the other hand, cases included in the low class of economic resilience showed high structural heterogeneity, highlighting the importance of applying specialized policies for promoting countries of low economic resilience. In terms of geography, the analysis filled the literature gap by shaping the worldwide map of economic resilience and revealed a pattern of a geographical duality between countries of stably high and stably medium to good economic resilience, along with centrifugal patterns of commute performance of economic resilience located in America, South Africa, and Europe, loosely suggesting markets of unsteady economic interaction. In terms of time, the analysis detected a relationship between diachronically good performance in economic resilience and geographical distance from the shocks' origin, although Asia and Europe were exemptions also benefited by internal crises. In the long run, Europe and Oceania we found the last in their economic resilience performance, America appeared to perform better in engineering resilience, while Africa and Asia in ecological and evolutionary resilience. The analysis also showed that Europe was most affected by the 2008 crisis, Africa, Asia, North America, South America, and Oceania, accrediting current literature in economic resilience that mainly focuses on the 2008 crisis in European Union. Finally, the analysis supported shaping the hypothesis that political instability is a main determinant of low economic resilience, addressing avenues of further research, along with these directions for further application of the proposed index.

## APPENDIX
### A.1. The 3D Resilience Index

In the first step, we develop a three-dimensional (3D) index of economic resilience, building on the context of capturing changes due to a shock (Martin and Sunley 2014; Psycharis et al., 2014; Kitsos and Bishop, 2018). The proposed index measures the engineering aspect of resilience in one dimension, the ecological aspect in a second dimension, and the evolutionary aspect in a third dimension. To incorporate in common all these three dimensions, we introduce the economic resilience index $I_R$ in a vector form, expressed by the mathematical formula:

$$I_R = (R_{en}, R_{ec}, R_{ev}) \quad (A1),$$

where $R_{en}$ represents the engineering, $R_{ec}$ the ecological, and $R_{ev}$ the evolutionary component. Under the relevant literature (Martin and Sunley, 2014; Kitsos and Bishop, 2018; Sdrolias et al., 2022), a critical choice for defining each component of the proposed vector index is to set a meaningful time reference for the conceptualization and measurement of economic resilience. Towards this direction, we define a variable $X$ as a proxy of a system's economic attribute, on which we can measure economic resilience, and, for a well-defined context, we claim variable $X$ to be a function of time $X=X(t)$. For the sake of intuition but without affecting the generality in definitions, we consider a discrete-time interval $t=1,2,\ldots,n$. Within this context, when a shock applies to a system at time $t_k$, it divides the time-series $t_X=\{t(x_1), t(x_2),\ldots, t(x_n)\}$ of variable's $X$ evolution into two periods: one before the shock $t_{Xk(-)}=\{t(x_1), t(x_2),\ldots, t(x_k)\}$, and a second one after the shock $t_{Xk(+)}=\{t(x_{k+1}), t(x_{k+2}),\ldots, t(x_n)\}$, as it is shown in Fig.A1. For the sake of simplicity, we name the first period as reference period ($R$) and the second one as the performance period (P). The names of these periods are descriptive because the first one defines a reference to the system's state of functionality before the shock (what was considered as normal functionality before the shock applies), while the second one is the period after the shock where the system's performance is submitted to examination whether it recovers or moves to a new state of functionality. Within each of these two periods, we can further define respectively two reference points: one in the reference period expressing the level ($c_R$) of the system's pre-shock performance (can be average, max, or other) and a second one in the performance period expressing the level ($M_R$) of the system's maximum performance after the shock.

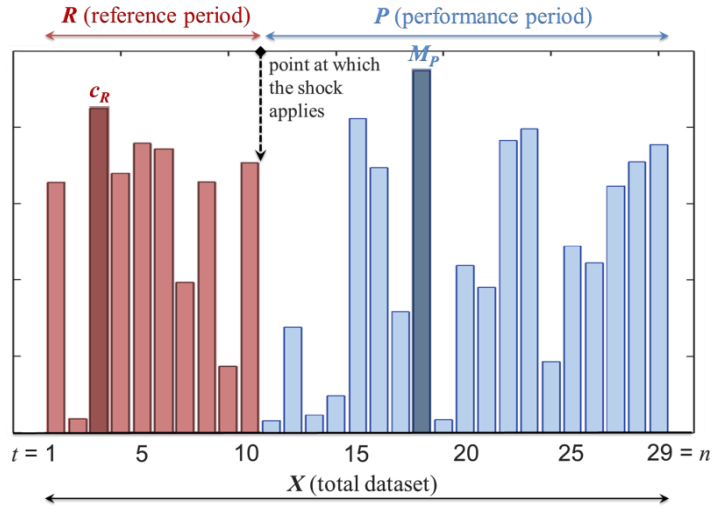

**Fig.A1.** A shock divides the time evolution of a proxy attribute X into a reference (R) and a performance (P) period, where we can define their concordant levels ($c_R$, $M_R$).

Within the context of the time reference illustrated in Fig.A1, we can define the mathematical expressions of the engineering ($R_{en}$), ecological ($R_{ec}$), and evolutionary ($R_{ev}$) components of the resilience index as follows:

■ *Engineering component ($R_{en}$):* The engineering indicator ($R_{en}$) measures the recovery speed (Fig.A2) of a system due to a shock, and is defined by the formula:

$$R_{en} = \frac{\log\left(\frac{n - t(c_R)}{t(x_{i \in P} \geq c_R) - t(c_R)}\right)}{\log\left(n - t(c_R)\right)} \quad (A2),$$

where $n$ is the length (number of elements) of the dataset X; $t(x_i)$ is the time point for attribute score $X = x_i$; R, P are the reference and performance periods; $c_R$ is the level of functionality in the reference period, defined by the maximum $c_R = \max\{X_R\}$; and $M_P$ is the level of functionality in the performance period, defined by the maximum $M_P = \max\{X_P\}$.

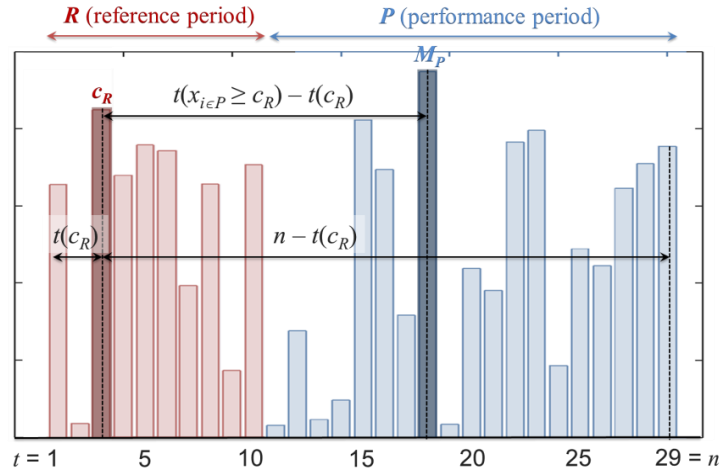

**Fig.A2.** Diagram illustrating the elements participating in the definition of the engineering (time recovery or horizontal shift) component of the 3D resilience index

The engineering component is defined along the horizontal axis and measures the time interval at which the system recovers to its previous level of functionality ($x_{i\in P} \geq c_R$), a fact that allows naming it also as a horizontal (shift) component. In particular, $R_{en}$ is defined based on the inverse proportion of (*i*) the recovery time ($t(x_{i\in P} \geq c_R) - t(c_R)$) to (*ii*) the total time after the shock ($n - t(c_R)$), so that to yield outcomes closer to the total time $n - t(c_R)$, for smaller recovery times, and equal to one, if the system does not succeed to recovery. To normalize the range [1, $n - t(c_R)$] into a typical indicator's interval [0,1], we, first, compute the logarithmic returns to gain a zero for the lower bound [log(1), $\log(n - t(c_R))$], and, secondly, we divide the converted interval with the new (logarithmic) upper bound to gain the one [$\frac{0}{\log(n-t(c_R))}, \frac{\log(n-t(c_R))}{\log(n-t(c_R))} = 1$]. According to this rationale, we divide with $\log(n - t(c_R))$ in the mathematical formula of $R_{en}$. Therefore, the mathematical expression of the engineering component yields: (*i*) zeros ($R_{en}$=0) in cases of no recovery, shifting the system to a new state of (either higher or lower) functionality; (*ii*) ones ($R_{en}$=1) for cases of immediate recovery (one time step after the emergence of the shock); and (*iii*) positive values smaller than one (0<$R_{en}$<1) and proportional to the duration, in cases of recovery.

■ *Ecological component*: The ecological indicator ($R_{ec}$) measures the vertical shift (Fig.A3) of a system due to a shock, expressing the difference in the level of a system's functionality before and after the shock, defined by the formula:

$$R_{ec} = \left(\exp\left\{\frac{|M_P - c_R|}{\max\{|c_R|, |M_P|\}}\right\}\right)^{\operatorname{sgn}^*\{c_R - M_P\}} \quad (A3),$$

where $R$, $P$ are the reference and performance periods; $c_R$ is the maximum level of functionality in the reference period; $M_P$ is the maximum level of functionality in the performance period; $|M_P - c_R|$ is the absolute difference between the levels of functionality before and after the shock, and $\text{sgn}^*\{c_R - M_P\}$ is a sign (signum) function, adjusted to yield also a positive sign to zeros (0 ≡ +1).

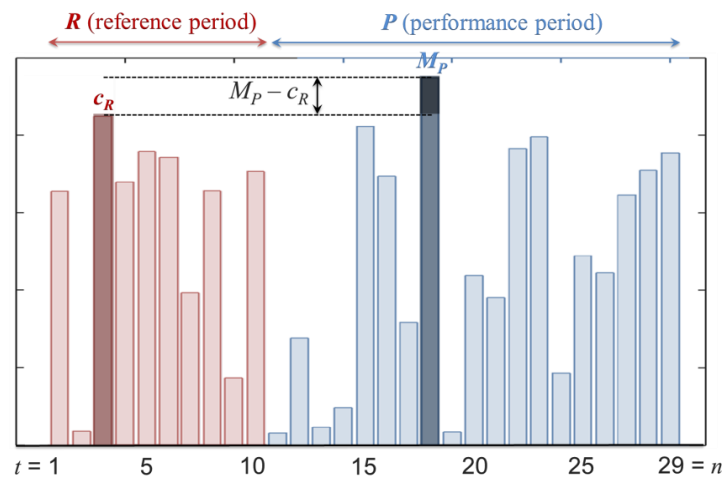

**Fig.A3.** Diagram illustrating the elements participating in the definition of the ecological (vertical shift) component of the 3D resilience index

The ecological component is defined along the vertical axis, a fact that allows naming it also as a vertical (shift) component. The $R_{ec}$ is defined on the rationale of computing the proportion of (*i*) the absolute difference in the level of functionality prior and after the shock ($|M_P - c_R|$) to (*ii*) the maximum between these two levels ($\max\{|c_R|, |M_P|\}$), yielding though zeros in cases system's recovery ($M_P = c_R$). In all other cases, the indicator yields positive outcomes smaller than one. Therefore, by applying an exponential transformation, we can convert the zero scores to ones and we can exponentially escalate all the other cases. In a further step, we apply an adjusted sign exponent to the expression $\exp\left\{\dfrac{|M_P - c_R|}{\max\{|c_R|, |M_P|\}}\right\}$, so that to inverse cases where the system moves to a state of lower functionality ($M_P < c_R$). Within this context, the ecological component yields scores: (*i*) positive and lower than one (0<$R_{ec}$<1) when the system shifts to a lower state of functionality; (*ii*) greater than one ($R_{ec}$>1) when it moves to states of higher functionality; and (*iii*) ones ($R_{ec}$=1) when it recovers to the previous state of functionality, implying an engineering performance.

  ■ *Evolutionary component*: The evolutionary indicator ($R_{ev}$) measures the adaptability of a system to continuously respond to changes induced by a shock, and is expressed in terms of after-shock variability (Fig.A4) according to the formula:

$$R_{ev} = \exp\left\{-\sum_{i \in P} \frac{c_R - x_i}{n(P) \cdot |c_R + M_P|}\right\} \quad (A4),$$

where $R$, $P$ are the reference and performance periods; $c_R$ is the maximum level of functionality in the reference period; $M_P$ is the maximum level of functionality in the performance period; $n(P)$ is the length (number of elements) of the performance period, and $c_R - x_i$ is the difference between the level of functionality in the reference period and the level of functionality in every step of the performance period.

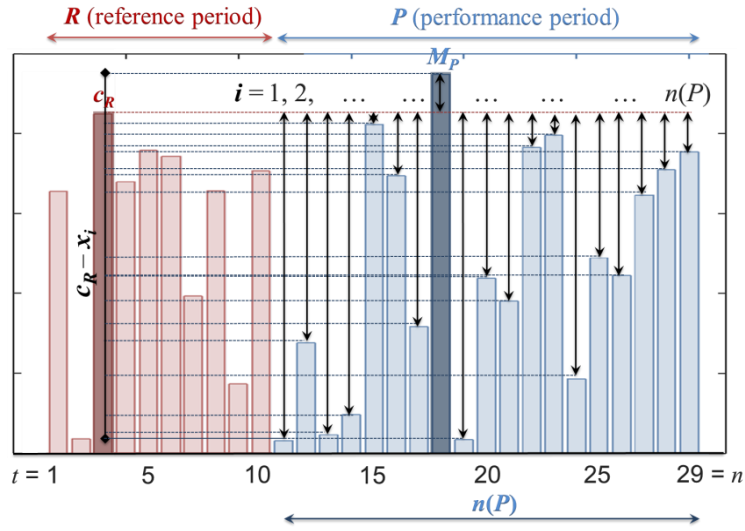

**Fig.A4.** Diagram illustrating the elements of the evolutionary (adaptability) component of the 3D resilience index

The evolutionary component is defined along both horizontal and vertical axes, on the rationale of measuring the variability of the performance period to the maximum level ($c_R$) of the reference period. Based on this conceptualization, we can also name this component as the adaptability component. According to its mathematical configuration, the evolutionary component yields scores: (*i*) equal to one (exp{0}=1) in case of a steady after-shock performance that equals to the pre-shock level of functionality ($c_R - x_i = 0$); (*ii*) positive and lower than one ($R_{ev} < 1$) when the system is described by variability below the pre-shock level of functionality ($c_R - x_i < 0$); (*iii*) significantly greater than one ($R_{ec} > 1$) when the system is described by a positively monotonic ecological process ($c_R - x_i > 0$); (*iv*) above one ($R_{ec} > 1$), when the average level of functionality in the performance period is higher than the level of the reference period ($\langle x_i \rangle > c_R$); and (*iv*) below one ($0 < R_{ec} < 1$), when the average level of functionality in the performance period is lower than the level of the reference period ($\langle x_i \rangle < c_R$).

- *The 1D economic resilience index*: For considering an aggregate, one-dimensional, version of the 3D economic resilience index ($I_R$), we introduce a scalar function $I_R : \mathbb{R} \to \mathbb{R}^3$ defined by the Euclidean norm (2-norm) of the indicator components, according to the formula:

$$I_R(\mathbf{I_R}, c_R) = \frac{\text{sgn}^*\{c_R - M_P\}}{\sqrt{3}} \cdot \|(R_{en}, R_{ec}, R_{ev})\|_2 \quad (5),$$

where $R$, $P$ are the reference and performance periods; $c_R$ is the level of functionality in the reference period; $M_P$ is the level of functionality in the performance period; $\text{sgn}^*\{c_R - M_P\}$ is the adjusted sign function (where $0 \equiv +1$); $R_{en}$, $R_{ec}$, and $R_{ev}$ are respectively the engineering, ecological, and evolutionary component of $I_R$; and $\|\cdot\|_2$ is the Euclidean (2-norm) norm. Intuitively, the factor $\frac{\|(R_{en}, R_{ec}, R_{ev})\|_2}{\sqrt{3}}$ in the formula of $I_R$ can perform as an "averaging" operator on the vector components ($R_{en}, R_{ec}, R_{ev}$) of the Euclidean norm, and it can thus evenly distribute the contribution of each $I_R$ component to the configuration of the numerical value of the 1D index of economic resilience. Finally, we incorporate in the formula of $I_R$ a coefficient defined by the adjusted sign function (yielding a positive sign also for zeros, $0 \equiv +1$), to regain the negative engineering information (the sign of $R_{en}$) that is lost due to the norm operation (producing only positive outputs). Within this context, (*i*) negative values of the 1D economic resilience index ($I_R<0$) indicate an overall deficient resilience profile described by a negative ecological performance (a shift of the system to a state of lower functionality), with simultaneous lack of engineering and perhaps (expected) deficient evolutionary performance; (*ii*) values greater than one ($I_R>1$) illustrate an overall high resilience profile described by an engineering performance of high recovery speed, with simultaneous positive ecological (a shift of the system to a state of lower functionality) and relatively high evolutionary performance; and (*iii*) intermediate values ($0<R_{en}<1$) illustrate cases of good or satisfactory overall resilience profile, which however is expected to be deficient in one or (maybe) two of the separate (engineering, ecological, evolutionary) components.

### A.2. Selection of Reference Crises (1970-2020)

**Table A1**
List of major modern financial crises (1970-2020)

| Reference Year | Crisis | Description / Basic Info | Geographical Scale of Impact | Reference |
|---|---|---|---|---|
| 1973 | Oil crisis | Began in October 1973 due to an oil embargo proclaimed by the members of the Organization of Arab Petroleum Exporting Countries. The embargo decreased the oil prices, causing a stock market crash and | Global | Akins (1973); Issawi (1978); Mitchell (2010) |

| Reference Year | Crisis | Description / Basic Info | Geographical Scale of Impact | Reference |
|---|---|---|---|---|
| | | secondary banking crisis. | | |
| 1980-1982 | Latin American debt crisis | Begun in the early 80s when the debt of Latin American countries overcame their repaying ability. The Latin American debt crisis caused economic recession (about 20-40%), unemployment, inequalities increase, and poverty. | International | Wiesner (1985); Sims and Romero (2013) |
| 1987 | Black Monday | The largest one-day percentage decline in stock market history, causing a sharp decline (from 20-50%) to 23 major world markets. | Global | Markham and Stephanz (1987); Schaede (1991); Metz (2003) |
| 1988 | Norwegian banking crisis | Started in 1988, in Norway, due to financial deregulation on lending rates' caps, which led banks to engage in more risky credit operations, resulting in a bank lending boom. It was one of the largest post-war financial crises in Europe that caused bank losses to many European countries. | National/ International | Englund (1999); Ongena et al. (2003) |
| 1989 | United States Savings & Loan crisis | A failure of almost one-third of Savings and Loan Associations in the USA, due to their social reform effort toward supporting house ownership of the working class. Caused considerable costs in the USA and the global economy. | National/ International | Pyle (1995); Curry and Shibut (2000); Barth et al. (2006) |
| 1990 | Japanese financial crisis (asset price bubble) | An economic bubble in Japan, where real estate and stock market prices inflated, causing overheat of economic activity, uncontrolled money supply, and credit expansion. It caused a decline (~15%) in commercial, residential, and industrial land prices in Japan. | National/ International | Hoshi and Kashyap (2004) |
| 1991 | Scandinavian banking crisis | Emerged as a housing bubble in Sweden and a systemic crisis of the entire financial sector in Finland, causing a severe credit crunch, bank insolvency, and significant losses of the national GDPs. | International/ Regional | Englund (1999); Ongena et al. (2003) |
| 1992-1993 | Black Wednesday | Started when the British Government withdrew the pound sterling from the European Exchange Rate Mechanism (ERM), after failing to comply with the lower currency ERM exchange limit. The crisis put the country into recession and damaged the credibility of the British economy. | International | Fratianni and Artis (1996); Soderlin (2000) |
| 1994 | The economic crisis in Mexico | A currency crisis due to a sudden devaluation of the Mexican currency against the U.S. dollar. It caused capital leaks, economic recession with a significant decline of GDP, prices increase, and unemployment and international impact. | National/ International | Carstens and Schwartz (1998); Davis-Friday and Gordon (2005) |
| 1997 | Asian Financial Crisis | It started with a financial collapse of Thailand's national currency and spread throughout the East and Southeast Asia countries, affecting 11 countries that raised fears of a worldwide economic meltdown. It | International | Maroney et al. (2004): Deesomsak et al. (2009) |

| Reference Year | Crisis | Description / Basic Info | Geographical Scale of Impact | Reference |
|---|---|---|---|---|
| | | had considerable macroeconomic effects, such as the reduction of currency power, stock market values, and other asset prices. | | |
| 1998 | Russian financial crisis | Emerged as a result of the impacts caused by the first war in Chechnya on the Russian economy, which resulted in declining productivity, national currency devaluation, and debt default. It had severe impacts on the national economy and the economies of many neighboring and related countries. | National/ International | Buchs (1999); Feridun (2004) |
| 1999 | Argentina economic crisis | A depression was affected by the Latin American debt crisis (the 1980s) and the Russian financial crisis (1998) and lasted almost four years. It caused unemployment, social and political riots, foreign debt default, a rise of alternative currencies, a shrinking of the economy (almost 30%), and poverty. | National/ International | Bebczuk and Galindo (2008) |
| 2000 | Turkish economic crisis | A financial crisis due to a high dependence on foreign investments and deficiency in their support of the national banking systems. It resulted in a stock market crash causing foreign divestment, corruption, and political instability. | National/ International | Alper (2001); Cizre and Yeldan, (2005) |
| 2001 | Bursting of the dot-com bubble | A stock market bubble caused by speculations of internet-related companies. Many companies crashed and others faced enormous decline. Stock markets faced great losses worldwide and the internet-related economic activity shrunk. | Global | Ljungqvist and Wilhelm (2003); Goodnight and Green (2010) |
| 2007-2008 | Worldwide financial crisis | A global financial crisis was mainly due to subprime lending to low-income homebuyers, financial institutions' excessive risk-taking, and the burst of the USA's asset property bubble. It caused worldwide (amongst others) great losses in asset property, disparity, unemployment, and poverty. | Global | French et al. (2009); Luchtenberg and Vu (2015) |

### A.3. Data and Variable Configuration

#### Table A2
List of socioeconomic variables* participating in the analysis

| Name | Description |
|---|---|
| Population | Total population, which counts all residents regardless of legal status or citizenship. Values are midyear estimates. |
| Urban Population | People living in urban areas. It is calculated using World Bank population estimates and urban ratios from the United Nations World Urbanization Prospects. |
| Population in the largest city | The percentage of a country's urban population living in that country's largest metropolitan area. |

| Name | Description |
|---|---|
| Labor force | Comprises people ages 15 and older, who supply labor for the production of goods and services during a specified period. It includes people who are currently employed and people who are unemployed but seeking work as well as first-time job-seekers. |
| Employment to population ratio | The proportion of a country's population that is employed. Ages 15 and older are generally considered the working-age population. |
| GDP per capita | Gross Domestic Product divided by midyear population. It is calculated without making deductions for depreciation of fabricated assets or depletion and degradation of natural resources (constant 2010 U.S. dollars). |
| GVA at basic prices | Gross value added at factor cost (formerly GDP at factor cost) is derived as the sum of the value-added in the agriculture, industry, and services sectors (constant 2010 U.S. dollars). |
| Prime sector GVA | Corresponds to ISIC divisions 1-5 and includes forestry, hunting, and fishing, as well as cultivation of crops and livestock production (constant 2010 U.S. dollars). |
| Industry GVA | It comprises value added in mining, manufacturing (also reported as a separate subgroup), construction, electricity, water, and gas (constant 2010 U.S. dollars). |
| Services GVA | Includes value added in wholesale and retail trade (including hotels and restaurants), transport, and government, financial, professional, and personal services such as education, health care, and real estate services. Also included are imputed bank service charges, import duties, and any statistical discrepancies noted by national compilers as well as discrepancies arising from rescaling (constant 2010 U.S. dollars). |
| Trade (% of GDP) | The sum of exports and imports of goods and services is measured as a share of gross domestic product. |
| Final consumption expenditure | The sum of household final consumption expenditure and general government final consumption expenditure (constant 2010 U.S. dollars). |
| Tax revenue (% of GDP) | Refers to compulsory transfers to the central government for public purposes (fines, penalties, and most social security contributions are excluded). Refunds and corrections of erroneously collected tax revenue are treated as negative revenue. |
| Total natural resources rents (% of GDP) | The sum of oil rents, natural gas rents, coal rents (hard and soft), mineral rents, and forest rents. |

Source: Worldbank (2021)